\documentclass[10pt,preprint]{aastex}
\usepackage{graphicx}
\usepackage{natbib}
\usepackage{amssymb}
\newcommand{\msun}{M$_{\odot}$}

\newcommand{\nsix}{NGC 6397}

\newcommand{\hst}{{\em HST}}
\newcommand{\acsbf}{$0.012\pm0.004$}
\newcommand{\wfpcbf}{$0.051\pm0.010$}
\newcommand{\gc}{globular cluster}
\newcommand{\blanki}{............................................}
\newcommand{\blankii}{.................} 
\begin{document}
\slugcomment{Draft to be submitted to AJ}
\title{The binary fraction in the globular cluster \nsix\altaffilmark{1}}
\author{
D.S.   Davis\altaffilmark{2},
H.B.   Richer\altaffilmark{2},
J.     Anderson\altaffilmark{3},
J.     Brewer\altaffilmark{2},
J.     Hurley\altaffilmark{4},
J.S    Kalirai\altaffilmark{5,6},
R.M.   Rich\altaffilmark{7},
P.B.   Stetson\altaffilmark{8}
}
\altaffiltext{1}{Based on observations with the NASA/ESA Hubble Space Telescope, obtained at the Space Telescope Science Institute, which is operated by the Association of Universities for Research in Astronomy, Inc., under NASA contract NAS5-26555. These observations are associated with proposals GO-10424.}
\altaffiltext{2}{Department of Physics \& Astronomy, University of British Columbia, 6224 Agricultural Rd., Vancouver, BC, V6T 3B4, Canada; sdavis@astro.ubc.ca}
\altaffiltext{3}{Department of Physics \& Astronomy, Rice University, Huston, Texas, USA}
\altaffiltext{4}{Center for Astrophysics \& Supercomputing, Swinburne University, Australia}
\altaffiltext{5}{University of California at Santa Cruz, Santa Cruz, CA, USA}
\altaffiltext{6}{Hubble Fellow}
\altaffiltext{7}{Division of Astronomy, University of California at Los Angeles, Los Angeles, USA}
\altaffiltext{8}{Herzberg Institute of Astrophysics, National Research Council of Canada, Victoria, Canada}

\begin{abstract}
Using Hubble Space Telescope (\hst) observations of the \gc\ \nsix, we constrain the cluster's binary fraction. The observations consist of two fields: the primary science field, a single ACS pointing centered approximately 5\arcmin\ from the cluster center; and the parallel field, a single WFPC2 field centered on the cluster center. Using the exquisite photometric precision of these observations, we determine the binary fraction in these regions of the cluster by examining stars lying off the main sequence.  The binary fraction is constrained  to be \acsbf\ in the ACS field, and  to be \wfpcbf\ in the WFPC field. N-body simulations by \cite{has07} suggest that the binary fraction remains nearly constant beyond the half-mass radius for the lifetime of the cluster. In the context of these simulations, our results suggest that \nsix\ had a primordial binary fraction of only $\sim$1\%.
\end{abstract}

\keywords{\gc s: individual (\nsix) --- stars:  Population II, Stellar Dynamics}

\section{Introduction}

Globular clusters (GCs) are fascinating dynamical testbeds. A key parameter to understanding the evolution of GCs is the binary fraction. The presence of primordial binaries can delay the onset of core collapse from on the order of $20$ half-mass relaxation times to over $150$ half-mass relaxation times \citep{hth06}. In a globular cluster, wide binaries with orbital speeds less than the velocity dispersion of the cluster will tend to be disrupted in three-body interactions. However, hard binaries will tend to have their orbits hardened by encounters with other cluster stars. The energy extracted from these systems is sufficient to support cluster cores from collapse \citep{hmg92}. The stable binary burning phase will last until the energy stored in primordial binaries has been exhausted. Core collapse then proceeds until the stellar densities increase to such an extent that binary formation via three-body interactions become important \citep{hth06}.

Additionally, the formation rates of stellar exotica, such as blue stragglers, cataclysmic variables, and low-mass X-ray binaries, are all strongly influenced by the binary fraction \citep{b95}. For instance, \cite{msf06} showed that while in the core of globular clusters blue stragglers form primarily by direct collisions, in the outskirts binary coalescence is the dominant formation mechanism. However, \cite{lsk07} recently found that in the cores of 57 globular clusters the frequency of blue stragglers did not scale with the collision rate, as would be expected if direct collisions were the dominant formation process. If binary coalescence is the dominant formation channel, we would expect the frequency of blue stragglers to scale tightly with the binary fraction. Despite the key role the binary fraction plays in our understanding of globular clusters, it remains a weakly constrained parameter in general.

The first searches for binaries in GCs examined stellar spectra for radial velocity variations \citep{gg79}. This technique relies on identifying individual binary systems, and hence, for a statistically significant sample, it is a time consuming method. Furthermore, searches of this type are only sensitive to certain orbital periods, inclinations, eccentricities, and mass ratios. Results for the whole cluster must be determined by extrapolating from a particular sample, making use of the uncertain distributions of the above mentioned parameters.

As large area CCDs and sophisticated photometric techniques became increasingly widespread, photometric accuracy allowed the search for binaries via their distinct colors and magnitudes \citep{rw91}. A main sequence star with a main sequence secondary will be  brighter and redder than a single star with a mass equal to that of the primary. The  maximum deviation from the main sequence is for equal-mass systems, which will have the same color as the primary, but will be brighter by $\sim0.75$ magnitudes. By modeling the distribution of stars in color-magnitude space as a function of the binary fraction, one can use the CMD as a statistical constraint on the binary fraction. Methods that make use of the CMD investigate the entire cluster in the field of view at once. This makes them intrinsically much less observationally intensive than methods that rely on studying individual systems one at a time, i.e., the detection of radial velocity variations or photometric variability.

While there seems to be a consensus that the present day binary fraction in globular clusters is clearly lower than in the disk of the Galaxy, the primordial binary fraction is still a matter of debate. Recently published papers on simulations of GCs include primordial binary fractions ranging from 5\% \citep{has07} to 100\% \citep{ibf05}. While it is unclear precisely how the binary fraction will evolve in the core of GCs, N-body simulations of \cite{has07} show that the binary fraction beyond the half-mass radius should remain stable. Thus, in the context of these simulations, a measurement of the present day binary fraction beyond the half-mass radius will constrain the cluster's primordial binary fraction. Here, we present a relatively precise determination of the binary fraction both within and outside the half-mass radius.

\section{Observations}
The observations reported in this paper were taken as part of GO-10424 \citep{rab06}. With the primary science goals of: (1) observation deeper than the theoretical termination of the main sequence in a globular cluster, (2) the first clear observation of the collision-induced absorption ``blue hook'' feature of the  white dwarf cooling sequence, and (3) the determination of the white dwarf cooling age of the cluster,  126 orbits were dedicated to imaging one ACS field 3\farcm17 E and 3\farcm93 S of the cluster core. For a sense of scale, R$_{\rm hm}$, the half-mass radius of the cluster, is 2\farcm33 \citep{har96}. The field was chosen for the wealth of archival data in that location. Archival WFPC2 images were taken in this field in 1994 and 1997, giving us a baseline of over a decade for the measurement of proper motions. The proper-motion cleaned CMD was critical for several projects. Unfortunately, the archival data only covered approximately 60\% of the ACS field, effectively reducing the size of the observed field. For the remainder of this paper we will refer to the overlap region between the archival WFPC2 images and the new ACS image as the ``outer field''.

\begin{figure*}
\plotone{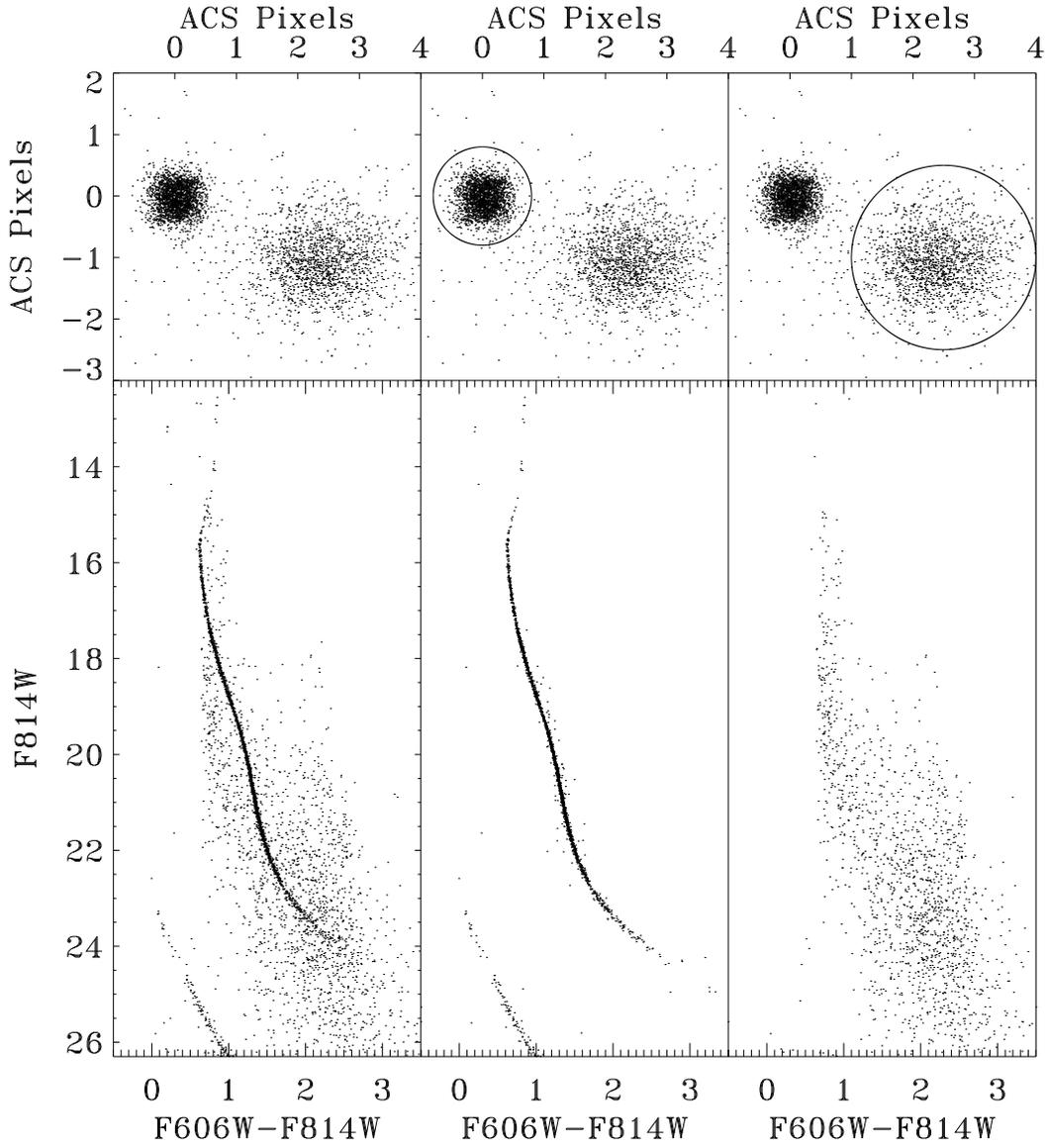}
\caption{The top panels show the observed proper motions  of the stars in the ACS field, while the bottom panels show color-magnitude diagrams for the same stars. The left panels show all the data. In the upper middle panel the cluster population is circled, and the lower middle panel shows the CMD of the circled population only. In the right panel, the field population is circled, and again in the lower right plot the CMD of the circled population in shown.\label{acscmd.fig}}
\end{figure*}

During this campaign, each orbit was divided into three exposures, with the first and last exposure of each orbit taken in F814W, and the middle exposure (with the darkest sky value) taken in F606W. In several orbits, short exposures were taken to measure bright stars that were saturated on the longer exposures. The total exposure time was 179.7 ks in F814W and 93.4 ks in F606W. The data were reduced using a novel technique designed to find and measure the faintest possible stars in a field where there are also very bright stars (for full details see Anderson et al. 2008). Sources are detectable to a depth of ${\rm F814W}\simeq30$, with the $50\%$ completeness limit being ${\rm F814W}\simeq28$. The proper motion measurements are severely limited by the archival WFPC2 data. Proper motion measurements are only reliable to a depth of ${\rm F814W}\simeq26.3$. This is {\em not} an impediment for this project, as we are concerned with only main sequence stars, the vast majority of which are brighter than ${\rm F814W}=25$. Figure \ref{acscmd.fig} shows the color-magnitude diagram (CMD) and the proper motions for the outer field, and demonstrates the utility of  proper motions to isolate the cluster. For the remainder of this paper, we will restrict ourselves to studying a section of color-magnitude space covering $16<{\rm F814W}<24$ and $0.6<{\rm F606W}-{\rm F814W}<2.4$ in this field.

During the primary target observations, the roll angle of the telescope was controlled in order to simultaneously obtain WFPC2 images centered on the core of the cluster. For the WFPC2, each orbit was also divided into three exposures. In addition to one in F814W and one in F606W, one exposure in F336W was obtained. Similar to the outer field, there were archival WFPC2 images in this field. Archival images obtained in 1996, 1999, and 2001 in both F606W and F814W enabled the measurement of proper motions over $\sim 75\%$ of the field. The overlap between the current WFPC2 images and the archival WFPC2 images will henceforth be referred to as the ``inner field''.

\begin{figure*}
\plotone{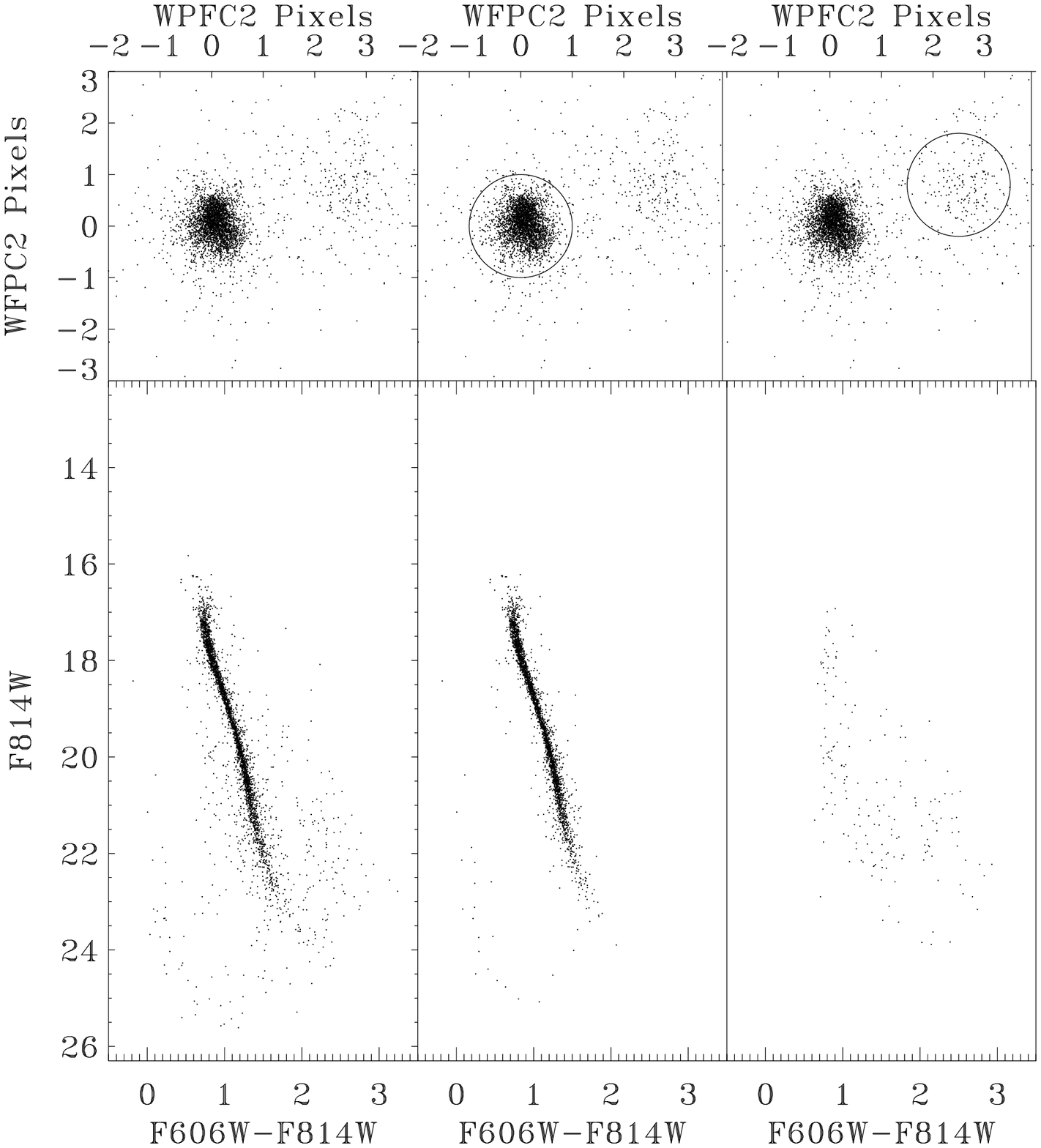}
\caption{The CMDs and proper motion diagram of the WFPC2 field. The various panels are analogous to Figure \ref{acscmd.fig}. \label{wfpccmd.fig}}
\end{figure*}

These images were reduced with standard DAOPHOT/Allstar \citep{ste87} techniques. The observations in the inner field had several limitations compared with the outer field. The stellar density in the inner field is significantly greater than that in the outer field, causing much more scattered light. Furthermore, the larger pixel size and lower sensitivity of the WFPC2 camera prevented the detection of sources fainter than ${\rm F606W}\simeq27$ and ${\rm F814W}\simeq26$. The degraded photometry of the measured stars resulted in a significantly broadened  main sequence compared with the outer field. Finally, short exposures were not taken, and hence all stars brighter than ${\rm F814W}\simeq17$ are saturated. The CMDs and proper motions of the inner field are shown in Figure \ref{wfpccmd.fig}. Because of the inferior quality of the photometry and astrometry of the inner field compared with the outer field, we restrict our study in this field to a smaller area of color-magnitude space: from $17<{\rm F814W}<22$ and $0.7<{\rm F606W}-{\rm F814W}<1.6$. For the remainder of the paper, figures will be shown for the outer field only, except when a difference between the fields is of particular interest.


\subsection{Optical binaries and field contamination}
In our proper-motion cleaned sample there are two sources that could contribute ``false'' binaries: optical binaries, and field stars that share the mean proper motion of the cluster. We now show that these sources of false signals will make only a minor contribution to the observed binary fraction.

\begin{figure*}
\plotone{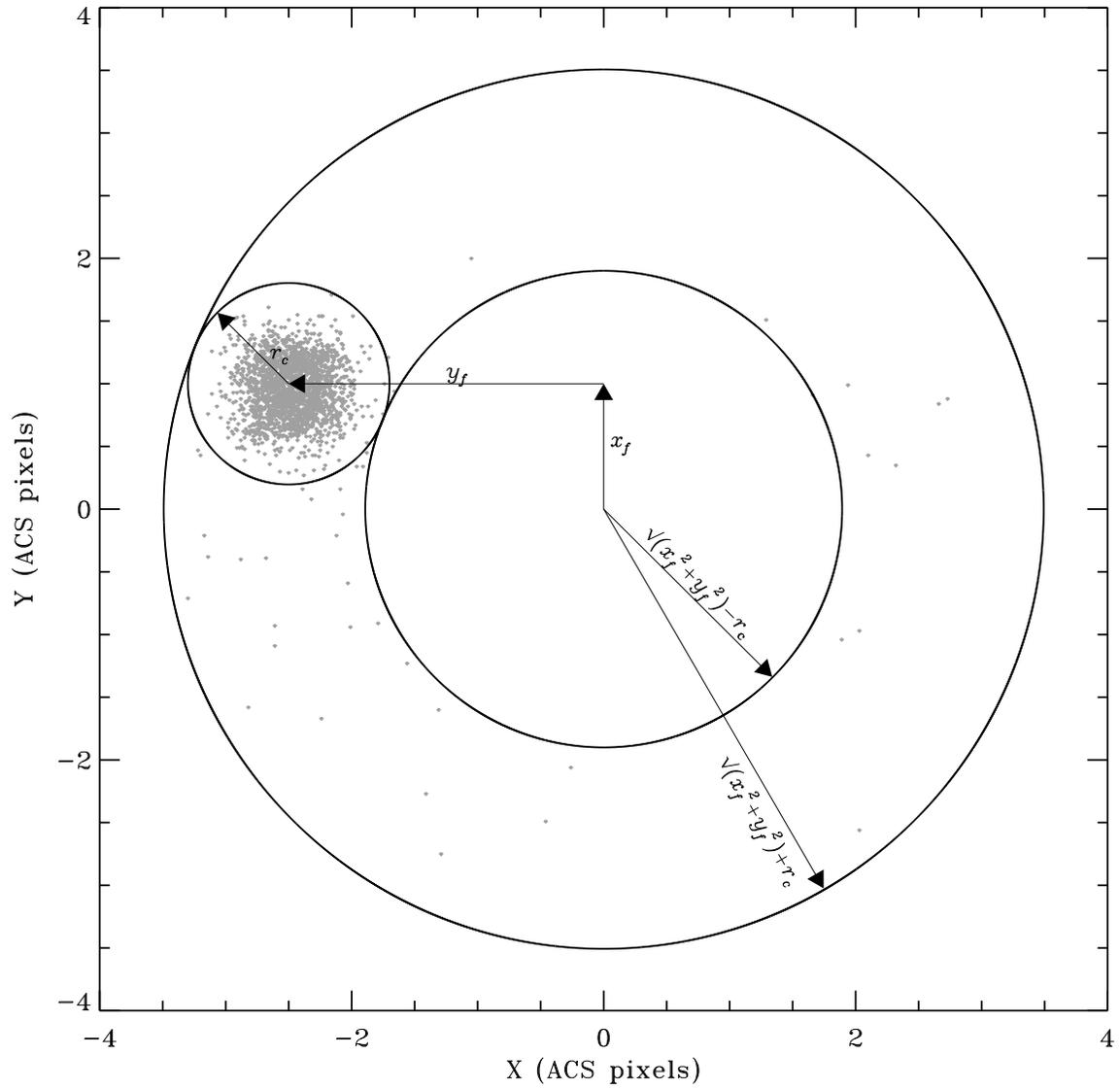}
\caption{The setup to calculate the contamination from field stars that share the cluster proper motion. \label{setup.fig}}
\end{figure*}

To address field star contamination we examine the proper motion diagrams using the setup shown in Figure  \ref{setup.fig}. Our criterion for classifying a star as a cluster member was that the magnitude of its proper motion relative to the mean motion of the cluster was less than a certain radius, $\left|\mu\right|<r_c$. The mean proper motion of the field was offset from the cluster by several pixels in the $x$ and $y$ directions, namely $x_f$ and $y_f$ respectively. In order to calculate the contamination from the field, the proper motions were re-centered on the field population. This change of coordinates modifies the cluster selection criterion for cluster stars. Cluster members now have the property that:
\begin{equation}
\sqrt{x_f^2+y_f^2}-r_c<\left|\mu\right|<\sqrt{x_f^2+y_f^2}+r_c.
\end{equation}
Selecting only stars in this annulus of proper motion space, we calculate the angle, $\theta$, of the proper motion vector for each star, with $\theta=0$ directed to the mean cluster proper motion. Stars in this annulus should come from both the cluster and the field populations. While the field population should be distributed uniformly with angle, the cluster population should be strongly concentrated on $\theta=0$.  This distribution is modeled as a Gaussian, representing the cluster, and a constant, representing the field population.  The number of contaminating stars is now estimated as the product of the total number of field stars in the annulus and the fraction of the annulus that is occupied by the cluster. The best-fit parameters from our two component model suggest that there are 95 field stars in the annulus for the outer field, implying approximately 7 field stars should have the same proper motion as the cluster.

However, field stars are only likely to have an effect on the binary fraction if they lie close to the binary sequence. Only $\sim10$\% of the field stars lie within the binary sequence in our color-magnitude space. We therefore expect only approximately one field star to contaminate the binary sequence.  There are $\sim 2.1\times10^3$ proper motion-selected cluster stars in the outer field. Given this number of cluster stars, one contaminating star will have a negligible effect on the measured binary fraction. A similar analysis was performed on the inner field. We expect the inner field CMD to be contaminated by $\sim8$ stars, but again, less than one star will lie on the binary sequence.

In order to get a sense of the contamination from the chance alignment of stars  we perform a rough calculation to assess the crowding of the field. The ACS camera consists of  $2\times2048\times4096$ pixels for a total of $16.8\times10^6$ pixels. We detected 8600 stars over the entire field, which implies we have one star for every 1950 pixels, or equivalently,  one star per 44 square pixels. Our detection method requires that a source must be the brightest source within 7.5 pixels, and thus each star will occupy roughly 177 pixels. This implies that our field could have had an order of magnitude more stars before being confusion limited, and so we expect the number of optical binaries to be relatively small. Because of this, we can treat the observed stellar number counts as a relatively good approximation of the true underlying number counts.

We next proceed to quantify the filling fraction as a function of magnitude, $f(m)$. This function quantifies, for a given magnitude, what fraction of the CCD is occupied by stars of that magnitude. For this study, it does not matter if two field stars overlap; these systems will be removed by their proper motions. However, two cluster stars that overlap, or a cluster and field star that overlap in the current epoch data will not be removed by proper motion measurement. We therefore calculate the filling fraction for cluster stars only. Although we require a star to be the brightest within 7.5 pixels in order to be detected, another star will interfere with the measurement of the magnitude of a star only if it lies within the FWHM of its PSF. Each star is therefore assumed to occupy $\pi r^2$ pixels, with $r=2$.

Only if two stars are of similar magnitude will their superposition affect the measured magnitude of the brighter star. From the width of our main sequence, we estimate that an increase in flux less than  $\sim10$\% would be lost in the general photometric scatter. Accordingly, for each magnitude, $m$, we calculate the number of stars with magnitudes between $m$ and $m_1$, where $m_1=m+2.5$. We call this quantity $n(m)$.

The number of contaminating stars as a function of magnitude, $c(m)$, is simply the product of the two previously introduced quantities, $c(m)=f(m)n(m)$. The total number of contaminating stars is then the integral of $c(m)$ over the appropriate range of magnitude. The total contamination from optical binaries in the ACS field is expected to be three stars, corresponding to an overestimation of the binary fraction of $\sim0.15$\%. Due to increased crowding, the optical binary contamination in the inner field is expected to be more serious. We predict roughly 25 optical binaries in the inner field, corresponding to an increased binary fraction of $\sim0.8$\%.

\section{The model}
\subsection{The single-star sequence}
With a small number of simple assumptions one can create a model of the distribution of single stars in a CMD. One only needs an isochrone that follows the main sequence ridge line (MSRL), a model for the photometric error as a function of magnitude, and a model of the mass function. These can all be determined from the observed stars.

\begin{figure*}
\plotone{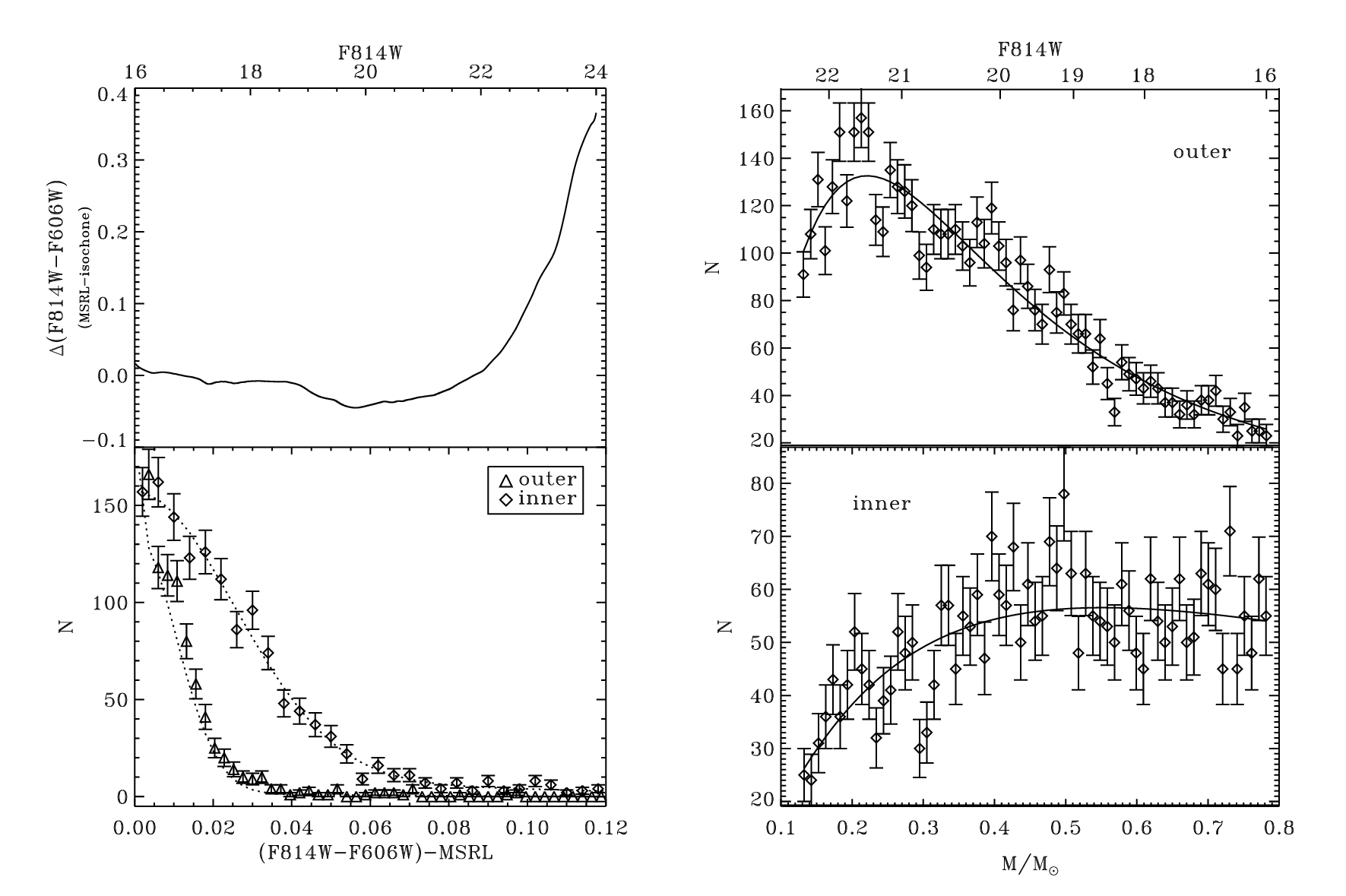}
\caption{The ingredients for simulating a single star sequence. {\bf \em Top left:} The difference between the colors of the empirically derived MSRL and the theoretical isochrone. The difference is very small brighter than F814W=22, and reaches a maximum deviation of $\sim0.35$ mags at F814W=24.
{\bf \em  Bottom left:} The blue-side scatter of (F606W$-$F814W)$-$MSRL, i.e., a ``straightened'' CMD, showing our observed photometric error. The fit of a model consisting of three superimposed Gaussians is shown with a dotted line.
{\bf \em  Top right:} The mass function of the outer field, with the log-normal fit shown as a line.
{\bf  \em Bottom right:} As for the top right, but for the inner field.
\label{mod.fig}}
\end{figure*}

In order to derive a relation between stellar mass and location on the MSRL,  we used stellar isochrones by \cite{dcj07}. Shown in Figure \ref{mod.fig} (top left panel) is the difference in color between the theoretical isochrones and the MSRL. For magnitudes brighter than ${\rm F814W}\simeq22$ the discrepancy between the MSRL and the isochrone is less than 0.03 magnitudes. However, the match between isochrone and MSRL degrades for stars fainter than this. This is most likely due to missing stellar atmosphere opacities at low temperatures. In order to correct the discrepancy, we assumed that the isochrone described an accurate relation between stellar mass and F814W. The mismatch between the isochrone and the MSRL only becomes important for masses less than $\sim0.17$ \msun. Due to the small difference in mass between this point and the hydrogen burning limit, the mass assigned to a particular location on the MSRL would not have been affected if it was assumed that the error was due entirely to the F814W magnitudes, and the F606W magnitudes were correct. The F606W values of the isochrone were adjusted to accord with the MSRL. The maximum deviation of the isochrone from the MSRL in the region we are examining is $\sim0.35$ mags. The match of these isochrones to our data is as good as with any other publicly available set.

The observed width of the main sequence is a combination of both intrinsic effects, such as differential reddening and minor differences in metalicity, and observational effects, such as scattered light and PSF variations. One can further subdivide the observational sources of error into those expected to be correlated and uncorrelated in the two observed bands. Sources of correlated errors, such as defocus, scattered light, and confusion, will have a smaller effect on the measured color of a star, and hence the broadening of the main sequence, than sources of uncorrelated errors, such as photon noise. For the purposes of our model, the source of the broadening is not important. We expect the photometric error to dominate the width of the main sequence of \nsix, and therefore we will refer to everything that broadens the main sequence as photometric error.

All main sequence-main sequence binaries will lie to the red side of the cluster main sequence, and thus scatter to the blue side should provide a reliable measure of the photometric error. The only sources of contamination here will be main sequence-white dwarf binaries and field stars that happen to share the mean proper motion of the cluster. Both of these are expected to have a negligible influence on the derived error. In order to model photometric error, we subtract the color of the MSRL from the measured color for each star. Then, for those scattering to the blue for the MSRL, we make a histogram of the number of stars and difference from the MSRL. The distribution was noted to have faint, but extended tails. Without in-depth analysis, the best we can do is to model the error as a superposition of Gaussians. Our error model is {\em not} intended to represent anything physical, but is rather an empirical fit. It was found that if we determined the  best-fit parameters for a superposition of two Gaussians, the tails were still not extended enough to accurately represent the observed scatter. A superposition of three Gaussians was found to model the observed blue-side scatter well. The blue-side scatter for both fields is shown in the lower left panel of Figure \ref{mod.fig}.  Although we need the error as a function of magnitude, the best-fit parameters were determined for the entire sample. This is because if we divide the sample into magnitude bins, the widths and heights of the best-fit Gaussians vary wildly, whereas we expect the underlying sources of error to vary smoothly and nearly monotonically with magnitude. In order give our error model some sensitivity to magnitude, the widths of the Gaussians were fixed, and their relative heights were allowed to vary as a function of magnitude.

While all clusters may be born with a ``universal'' initial mass function, dynamical evolution will alter the mass function and make it a function of position in the cluster. Furthermore, observational complications, such as incompleteness, make the determination of the ``true'' mass function difficult. As with the photometric error, for the purposes of this project we do not need to disentangle physical effects from observational effects; we are solely concerned with the {\em observed} mass function. As shown in Figure \ref{mod.fig} (right panels), both the fields were satisfactorily fit by a log-normal function:
\begin{equation}
{\rm N(M)}={\rm N}_0~\exp\left[\left(\frac{-\log_{10}({\rm M}/{\rm M}_0)}{\sqrt{2}\sigma}\right)^2\right].
\end{equation}
The best fit values for the outer (inner) field were: N$_0=132.6~(56.56)$, M$_0/$M$_\odot=0.22~(0.55)$, and $\sigma=0.30~(0.50)$ yielding a reduced $\chi^2=1.42~(1.25)$.

With the determination of the mass functions, photometric errors, and isochrones, we are now in a position to determine the probability distribution of the single star sequence in color-magnitude space, $F_s(c,m)$. We pixelize the color-magnitude plane finely enough that the minimum width of the main sequence is sampled by two pixels. This amounts to $450\times450$ pixels for the color-magnitude space in the outer field, and $200\times200$ pixels in the inner field. Over-sampling the color-magnitude plane does not inhibit the determination of the binary fraction, but adds computational time. Under sampling the color-magnitude plane will degrade the sensitivity of the experiments \citep{rw91}.

\begin{figure*}
\plotone{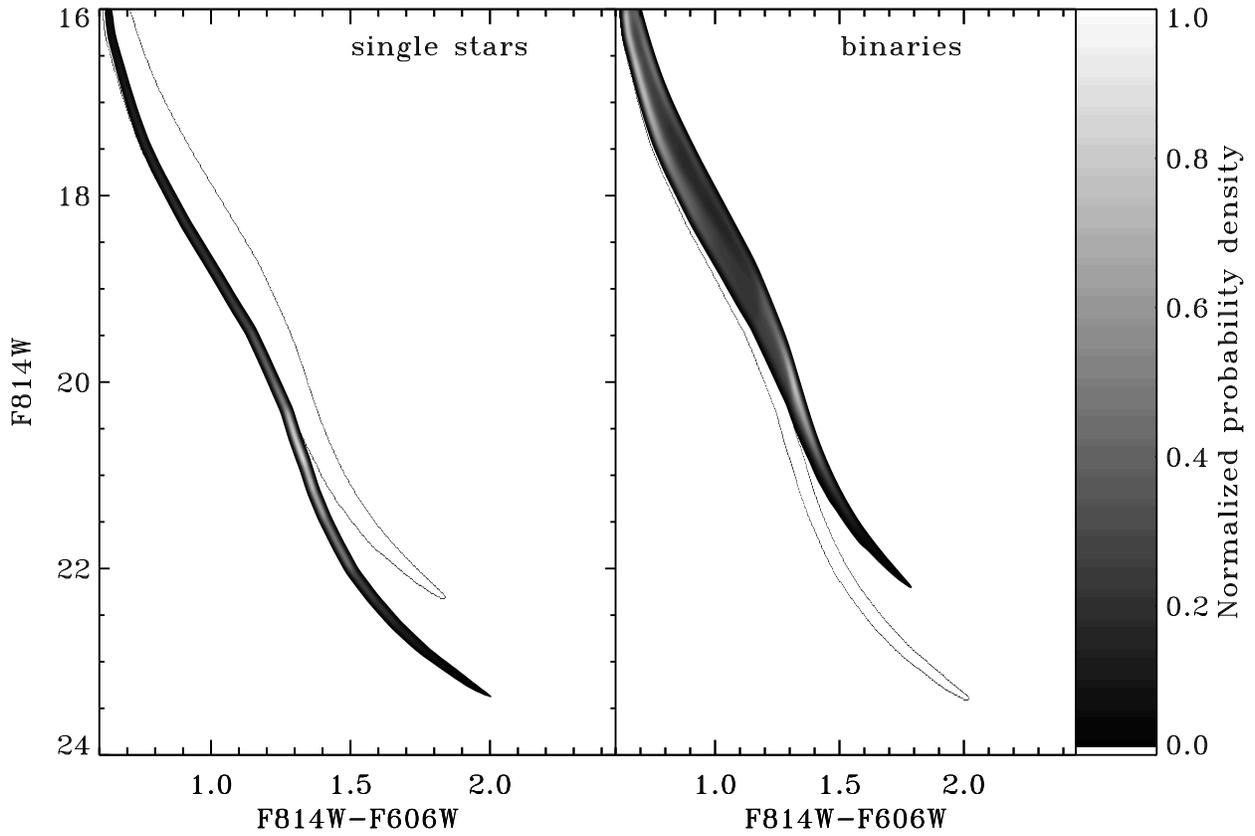}
\caption{{\bf \em Left:} The CMD showing the probability distribution of the single star sequence. The outlined area shows the ``footprint'' of the binary sequence, i.e., the area in which the probability of finding a star from the binary distribution is greater than 2\% of the peak value of the binary distribution. {\bf \em Right:} The CMD showing the probability distribution of the binary sequence. The outlined area shows the footprint of the single star sequence. \label{prob.fig}}
\end{figure*}

The isochrone is interpolated so the mass points are evenly spaced. We found we needed to have a mass interval of $10^{-4}$\msun\ in order to ensure points would fall in continuous pixels at the low-mass end of the main sequence.  In order to calculate $F_s$, the color and magnitude are calculated for each mass, and the value of the pixel corresponding to that color and magnitude is increased by the amount proportional to the value of the mass function for that mass. For example, from the upper right panel of Figure \ref{mod.fig}, one can see that the value of the color-magnitude pixel corresponding to a 0.2 \msun\ star would be increased by 130 units, while the pixel corresponding to a 0.7 \msun would be increased by 35 units. After the single star sequence has been added to the color-magnitude plane, it is convolved with the model for the photometric error. Finally, the resultant array, $F_s$, is normalized so that integrating over the entire color-magnitude space examined yields a probability of unity.  Figure \ref{prob.fig} (left panel) shows $F_s$ for the outer field.

\subsection{The binary sequence}

\begin{figure}
\plotone{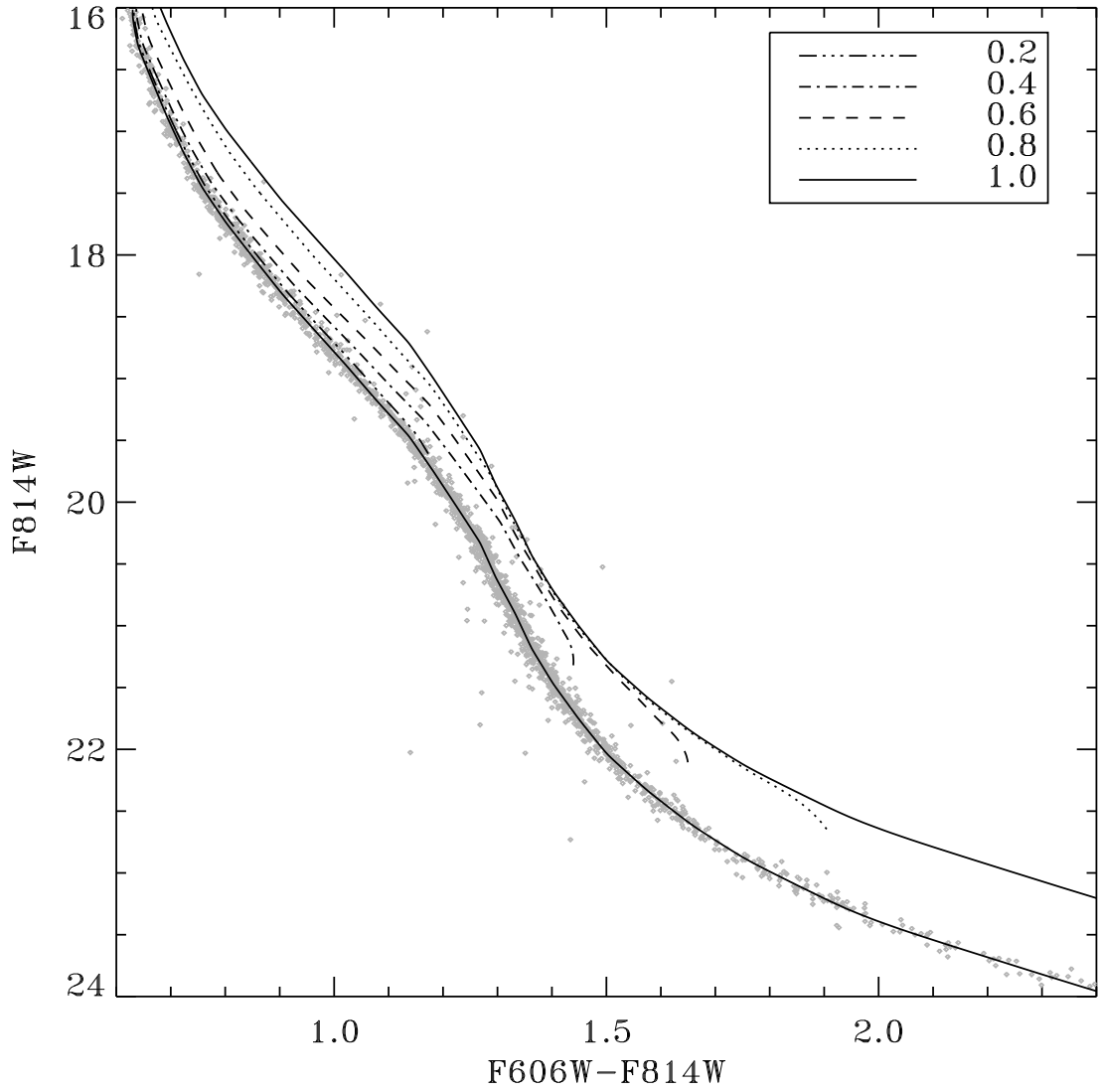}
\caption{The CMD of the outer field. Overlaid is the MSRL, as well as the expected ridge lines for binary star sequences of several mass ratios.\label{mr.fig}}
\end{figure}

A unresolved binary system will photometrically lie off the main sequence in the color-magnitude plane. The deviation from the MSRL is a function of the mass ratio, $q=M_2/M_1$. The deviation from the main sequence in magnitude increases monotonically, with the maximum deviation of $\sim0.75$ magnitudes for $q=1$. The deviation in color is always to the red, with a maximum deviation for $q\simeq0.8$ and no deviation for $q=0$ and $q=1$. Shown in Figure \ref{mr.fig} is the CMD of the outer field with the MSRL overlaid, and the deviation from the MSRL of binaries with various mass ratios. Mass ratios greater than $0.4$ are clearly distinguishable from the main sequence for most of the color-magnitude space that we are exploring.

The only further assumption that needs to be made to simulate the color-magnitude distribution of binary stars, $F_b(c,m)$, is the  distribution of $q$'s, referred to here as the $q$ function, as in analogy to the mass function. Both dynamics \citep[][and references therein]{gie05} and observations \citep{ps06,fss05} suggest that the $q$ function should favor mass ratios close to unity over extreme mass ratio objects.  \cite{ps06} suggest a population with $55 \%$ of the $q$'s drawn from a flat distribution, with the remainder drawn from a ``twin'' population (q$>$0.95). However, \cite{ps06} are exploring a very different regime in terms of mass ratio and primary masses than we are in \nsix; for all the systems included in their study $q>0.45$ and $M_1>7$ \msun. \cite{fss05} constrain the $q$ function with $M_1\gtrsim 1 $ \msun. While they do not provide a functional form for the $q$ function, they also find the distribution to be strongly peaked for $q$'s close to unity. The simulation in \cite{has07}  was well fit over the range $0.18<q<1$ by a power law, $N=N_0q^\alpha$, with $\alpha=1.7$. We are not suggesting that a power law has a physical basis; however, we require an analytical form in order to model the data.

We  modeled the $q$ function with a power law with  three characteristic values of the exponent: $\alpha=$ 0, 5/3, and 3. The method for determining the probability distribution of binaries in color-magnitude space, $F_b$, is very similar to that for $F_s$. Each star in the isochrone is chosen sequentially as the primary mass. Then, for all $q$'s that imply a secondary mass greater than the hydrogen-burning limit, the luminosities of the primary and secondaries are combined, and the magnitude and color of the system is calculated. The value in the pixel corresponding to that color and magnitude is then increased  according to the value determined from both the mass function (with respect to the primary) and the $q$ function. Similar to the single star sequence, after the probabilities for all binaries have been calculated, the composite probability is convolved with a model of the photometric error and normalized to unity. $F_b$ is shown in Figure \ref{prob.fig} (right panel).

\section{Results}
During this exercise, $F_s$ and $F_b$ need to be calculated only once for each value of $\alpha$. Then, for any given binary fraction, $f$, the probability of observing a star can be written as:
\begin{equation}
F_t(c,m\left|f\right.)=(1-f)F_s(c,m)+fF_b(c,m).\label{ft.equ}
\end{equation}
Our model of the observed color-magnitude diagram is therefore a family of two parameters: the binary fraction, $f$, explicitly through the dependence on Equation \ref{ft.equ}; and $\alpha$, implicitly through its influence on $F_b$.

\begin{figure*}
\plotone{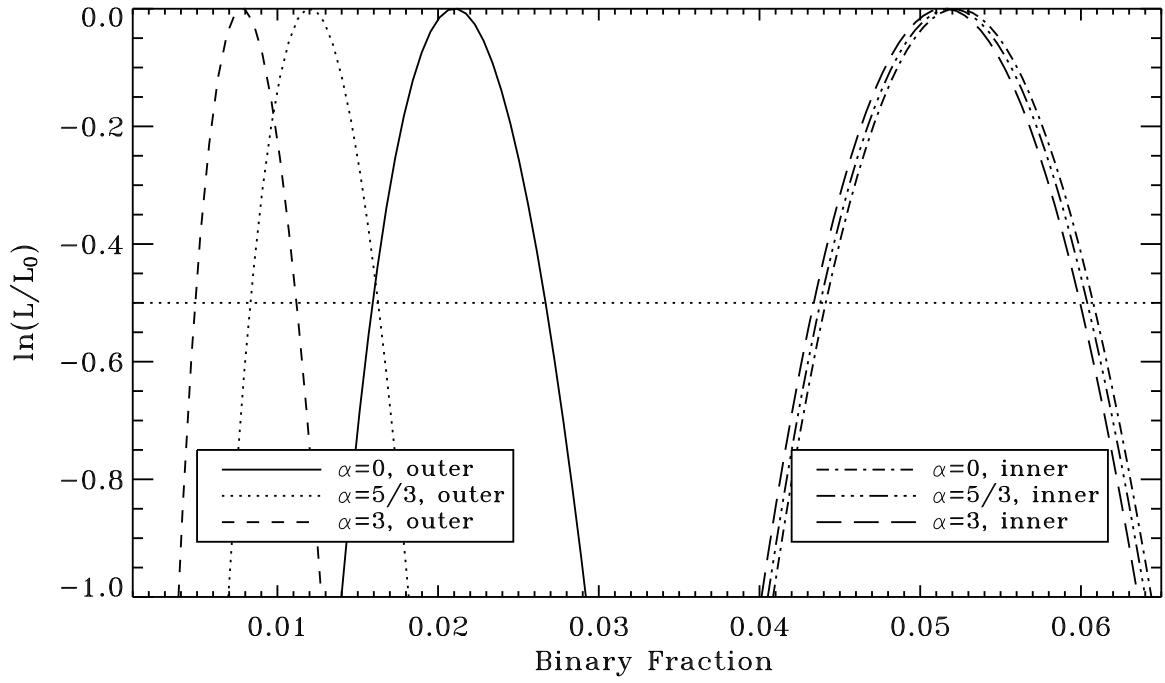}
\caption{The likelihoods of various models for the outer (left) and inner (right) fields.\label{results.fig}}
\end{figure*}

For a given $\alpha$, the best fitting $f$ is estimated by the maximum likelihood (ML) method. The probability of finding $n$ stars in a particular color-magnitude range is simply given by the Poisson distribution. For this study, the explicit form is:
\begin{equation}
P(c,m\left|\right.n)=F_t(c,m\left|\right.f)^ne^{F_t(c,m\left|\right.f)}/n!.
\end{equation}
The likelihood of the data given the model is then:
\begin{equation}
L(f)=\prod_{i=0}^N P_i\left(c,m\left|\right.n\right).
\end{equation}
The absolute value of $L(f)$ is meaningless; however, as $N$ becomes large, the value of $f$ that maximizes $L(f)$ will tend to the true value of $f$, namely $f_0$. As shown in \cite{rw91}, for one degree of freedom, the ``1-$\sigma$'' confidence in $f$ will be the value of $f$ such that ln$f$ drops from its peak value by 0.5. The likelihoods for the various models for the outer field and inner field are shown in Figure \ref{results.fig}. The results are relatively insensitive to the choice of $\alpha$---the difference between the most likely values is $\sim0.001$ and $\sim0.01$for the inner and outer fields respectively. The derived binary fraction in the outer field is $0.021 \pm 0.005$, $0.012 \pm 0.004$, and $0.008 \pm 0.003$ for $\alpha=0$, 5/3, and 3 respectively. Likewise, for the inner field, the derived binary fraction is $0.052 \pm 0.010$, $0.051 \pm 0.010$, and $0.051 \pm 0.010$ for $\alpha=0$, 5/3, and 3 respectively. Because of the observed $\alpha$ in the  models of \cite{has07}, we prefer the $\alpha=5/3$ value.

\begin{deluxetable}{llr|clr}
\tablecaption{Literature binary fraction constraints \label{prop.tab}}
\tablecolumns{6}
\startdata

\multicolumn{1}{c}{}& \multicolumn{2}{c}{inside half-mass radius} & & \multicolumn{2}{c}{outside half-mass radius}  \\
\multicolumn{1}{c}{}&\multicolumn{1}{c}{$f$}&\multicolumn{1}{c}{reference}& &\multicolumn{1}{c}{$f$}&\multicolumn{1}{c}{reference}\\
 \hline\hline
NGC 288....       & $0.15\pm0.05$            & ........\cite{bfm02}       & & $0.0^{+0.1}_{-0.0}$......& .........\cite{bfm02}          \\
..................& $>0.06$.......           & ............\cite{sbf07}   & & \blankii                 & \blanki                      \\
NGC 362....       & $0.21\pm0.06$            & ............\cite{fwm93}   & & \blankii                 & \blanki                      \\
NGC 2808..        & \blankii                 & \blanki                    & & $0.20\pm0.04$            & ............\cite{ala98}       \\
NGC 3201..        & \blankii                 & \blanki                    & & $<0.1$........           & ................\cite{cwf94}   \\
NGC 4590..        & $>0.09$.......           & ............\cite{sbf07}   & & \blankii                 & \blanki                      \\
NGC 5053..        & $>0.08$.......           & ............\cite{sbf07}   & & \blankii                 & \blanki                      \\
NGC 5466..        & $>0.08$.......           & ............\cite{sbf07}   & & \blankii                 & \blanki                      \\
NGC 5897..        & $>0.07$.......           & ............\cite{sbf07}   & & \blankii                 & \blanki                      \\
NGC 6101..        & $>0.09$.......           & ............\cite{sbf07}   & & \blankii                 & \blanki                      \\
NGC 6362..        & $>0.06$.......           & ............\cite{sbf07}   & & \blankii                 & \blanki                      \\
NGC 6397..        & $<0.07$.......           & .........\cite{cb02}       & & \blankii                 & \blanki                      \\
NGC 6723..        & $>0.06$.......           & ............\cite{sbf07}   & & \blankii                 & \blanki                      \\
NGC 6752..        & $0.27\pm0.12$            & \cite{rb97}                & &$0.02^{+0.16}_{-0.02}$... & \cite{rb97}                    \\
NGC 6792..        & \blankii                 & \blanki                    & &``low''........           & ............\cite{cpz07}       \\
NGC 6981..        & $>0.10$.......           & ............\cite{sbf07}   & & \blankii                 & \blanki                      \\
M3.............   & \blankii                 & \blanki                    & &``low''........           & .........\cite{gg79}           \\
..................& \blankii                 & \blanki                    & &$\sim0.04$.......         & ...............\cite{plh88}    \\
..................& $0.14\pm0.08$            & .........\cite{zb05}       & &$0.02\pm0.01$             & ..........\cite{zb05}          \\
M4.............   & $0.23^{+0.34}_{-0.23}$...& .........\cite{cf96}       & &$\sim0.02$.......         & ..............\cite{rfb04}     \\
M15...........    & $\sim0.07$.......        & .........\cite{gpw94}      & & \blankii                 & \blanki                      \\
M22...........    & \blankii                 & \blanki                    & &$0.03^{+0.16}_{-0.03}$... & ................\cite{cpm96}   \\
M30...........    & \blankii                 & \blanki                    & &$<0.05$.......            & ............\cite{ala98}       \\
M55...........    & $>0.06$.......           & ............\cite{sbf07}   & & \blankii                 & \blanki                      \\
M71...........    & $0.22^{+0.26}_{-0.12}$...& ............\cite{ym94}    & & \blankii                 & \blanki                      \\
M92...........    & \blankii                 & \blanki                    & &$0.00^{+0.03}_{-0.00}$... & ..................\cite{and97} \\
Arp 2.........    & $>0.08$.......           & ............\cite{sbf07}   & & \blankii                 & \blanki                      \\
Terzan 7....      & $>0.21$.......           & ............\cite{sbf07}   & & \blankii                 & \blanki                      \\
Palmoar 12        & $>0.18$.......           & ............\cite{sbf07}   & & \blankii                 & \blanki                      \\
Palmoar 13        & $0.30\pm0.04$            & ...............\cite{csb04}& & \blankii                 & \blanki                      \\
47 Tucane..       & $0.14\pm0.04$            & .............\cite{agb01}  & &$>0.05$.......            & \cite{dp95}                    \\
..................& \blankii                 & \blanki                    & &$\sim0.02$.......         & ..................\cite{and97} \\
$\omega$ Centauri & \blankii                 & \blanki                    & &$<0.05$.......            & ...............\cite{egs95}    \\
\enddata
\label{prev.tab}
\end{deluxetable}

\section{Conclusion}
This paper presents a precise determination of the binary fraction of \nsix. The binary fraction in the outer field (ranging from 1.3--2.8 R$_{\rm hm}$) is $0.012\pm0.004$, and $0.051\pm0.010$ in the inner field ($< 1$ R$_{\rm hm}$). The simulations of \cite{has07} suggest the present day binary fraction beyond the half-mass radius is very close to the primordial binary fraction, while the binary fraction in the central regions of the cluster is enhanced. In this context, our results suggest the primordial binary fraction of \nsix\ is $\sim$ 1\%.

Compiled in Table \ref{prev.tab} are literature values for previous attempts to constrain the binary fraction in \nsix\ and other Galactic GCs. The values determined interior to R$_{\rm hm}$ vary widely. This can be understood in the context of varying dynamical states. In sharp contrast to the values found in the core, the values determined beyond R$_{\rm hm}$, are generally quite low. Confounding factors obscure the interpretation of this result. For instance, \cite{bis06} suggest that the initial binary fraction may be a function of cluster metallicity. Additionally, \cite{lad06} find that binary fraction is a function of individual stellar spectral type, with $f$ peaking for G-type stars, and declining for stars of lower mass. While it would be naive to assume a univeral low binary fraction for all Galactic GCs, it seems clear that to construct accurate dynamical models of many clusters, a low primordial binary fraction must be used.

\acknowledgments
DSD would like to thank I.R.~King and J.V.~Wall for useful conversations and the UBC Messenger-UGF for funding. HBR is generally funded by  NSERC, but support for this project was also provided through UBC by the Vice President for Research, the Dean of Science, the NSERC Emergency Fund, and the Department of Physics and Astronomy. JA  and JSK  received support from NASA/HST through grant GO-10424. JSK is supported by NASA through Hubble Fellowship grant HF-01185.01-A, awarded by the Space Telescope Science Institute, which is operated by the Association of Universities for Research in Astronomy, Incorporated, under NASA contract NAS5-26555.

\clearpage
\bibliographystyle{apj}

\end{document}